\documentclass[aps,twocolumn,
superscriptaddress,
footinbib,
noeprint,
longbibliography,
prb]{revtex4-2}

\usepackage{amsmath}
\usepackage{listings}
\usepackage{graphicx}
\usepackage{dcolumn}
\usepackage{bm}
\usepackage{graphics}
\usepackage{bbold}
\usepackage{epsfig,color}
\usepackage{dsfont}
\usepackage{mathtools,amssymb}
\usepackage{braket}
\usepackage{stmaryrd}
\usepackage{bbm}

\usepackage{tikz}
\usetikzlibrary{arrows.meta,positioning}

\usepackage[breaklinks,colorlinks,bookmarks=false,citecolor=red,linkcolor=blue,urlcolor=magenta]{hyperref}

\begin{document}

\title{Probing the Spacetime Structure of Entanglement in Monitored Quantum Circuits with Graph Neural Networks}

\author{ Javad Vahedi}
\email[]{javahedi@gmail.com}
\affiliation{Department of Physics and Earth Sciences and
Department of Computer Science, Constructor University, Bremen 28759, Germany}

\author{Stefan Kettemann}
\email[]{skettemann@constructor.university}
\affiliation{Department of Physics and Earth Sciences and
Department of Computer Science, Constructor University, Bremen 28759, Germany}

\date{\today}

\begin{abstract}
Global entanglement in quantum many-body systems is inherently nonlocal, raising the question of whether it can be inferred from local observations. We investigate this problem in monitored quantum circuits, where projective measurements generate classical records distributed across spacetime. Using graph neural networks (GNNs), we represent individual quantum trajectories as directed spacetime graphs and reconstruct the half-chain entanglement entropy from local measurement data alone. Because information propagates through the network via local message passing, the architecture directly controls the spacetime region over which correlations can be aggregated. By systematically varying this accessible scale—through network depth and hierarchical spacetime coarse-graining—we probe how much measurement information is required to reconstruct global entanglement. We find that prediction accuracy improves as the accessible spacetime region grows and that results from different architectures collapse when expressed in terms of an effective spacetime scale combining depth and coarse-graining. These results demonstrate that the information required to reconstruct global entanglement is organized in spacetime scales and show that graph-based learning architectures provide a controlled operational framework for probing how global quantum correlations emerge from local measurement data.
\end{abstract}

\maketitle

\section{Introduction}

Understanding how global quantum correlations emerge from local
information is a central challenge in quantum many-body physics.
Entanglement entropy provides a quantitative measure of nonlocal
quantum correlations and plays a crucial role in quantum information
processing, thermalization, and emergent collective behavior
\cite{Emerson2003,Oliveira2007}. Because entanglement is inherently
nonlocal, determining it typically requires full knowledge of the
quantum state, which becomes exponentially costly for large systems.
This raises a fundamental question: to what extent can global
entanglement properties be inferred from local observations?

Recent developments in monitored quantum systems provide a natural
framework in which this question can be explored. In monitored quantum
circuits—unitary dynamics interspersed with stochastic projective
measurements—the measurement record generates a classical dataset
distributed across spacetime. Each measurement outcome provides local
information about the system at a particular qubit and time, while the
global entanglement structure of the underlying quantum state evolves
in response to both unitary dynamics and measurements. Such systems
have attracted considerable attention due to the discovery of
measurement-induced phase transitions (MIPTs), in which the
entanglement scaling changes from a volume law to an area law as the
measurement rate increases \cite{Li2018,Skinner2019,Chan2019,Gullans2019}.
Although this transition has been extensively studied, the broader
question of whether global entanglement observables can be reconstructed
directly from the local measurement record remains largely unexplored.

Machine learning provides a promising framework for addressing this
problem. Neural networks have recently been used to represent quantum
states, classify phases of matter, and detect phase transitions from
experimental or simulated data
\cite{Ohtsuki2016,Carleo2017,Carrasquilla2017,Mehta2019}. In the context of
monitored circuits, learning-based approaches have been used to
identify measurement-induced transitions or decode quantum
information from measurement outcomes
\cite{Dehghani2023,ElAyachi2024}. However, most architectures employed
in these studies—such as fully connected networks—do not explicitly
incorporate the spacetime structure of the measurement record. As a
result, they provide limited insight into how information propagates
across spacetime and how much of the measurement history is required
to reconstruct global observables.

In this work, we introduce a framework in which graph neural networks (GNNs) serve as controlled probes of the spacetime structure of entanglement in monitored quantum circuits~\cite{Scarselli2009,Hamilton2017}. Quantum trajectories naturally define directed spacetime graphs whose nodes correspond to qubit–time events and whose edges encode causal and entangling relations. Because GNNs operate through local message passing, information can propagate only along graph edges, such that the network architecture explicitly controls the spacetime region over which correlations are aggregated.

We exploit this property to systematically vary the accessible spacetime scale and investigate how much local measurement information is required to reconstruct global entanglement observables. We compare single-scale architectures, in which the receptive field grows linearly with network depth, with hierarchical architectures based on iterative $2\times2$ spacetime blocking that extend the effective spacetime scale through coarse-graining.

We find that the half-chain entanglement entropy of monitored quantum circuits can be accurately reconstructed from local measurement records when information is aggregated over sufficiently large spacetime regions. Prediction accuracy is governed primarily by the accessible spacetime scale: increasing this scale improves performance, and results from different architectures collapse when expressed in terms of a unified effective spacetime scale that combines depth and coarse-graining. These results demonstrate that the information required to reconstruct global entanglement is organized across spacetime scales and establish graph-based learning architectures as an operational framework for probing how global quantum correlations emerge from local measurement data.

\begin{figure*}
\centering
\includegraphics[width=\linewidth]{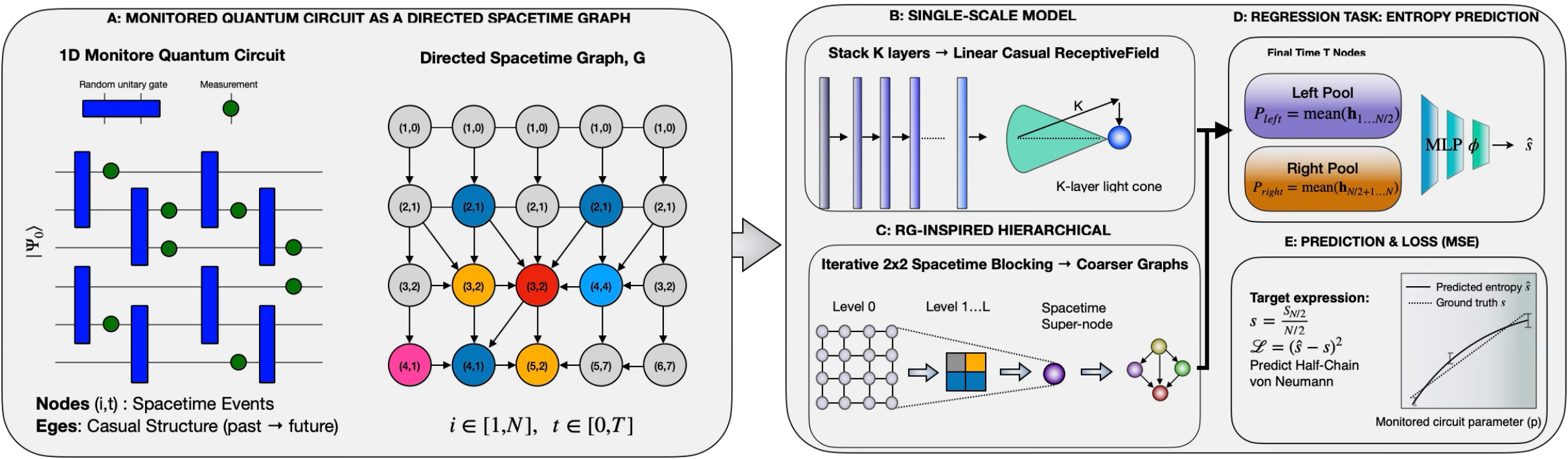}
\caption{
Overview of the learning framework for predicting entanglement in monitored quantum circuits. (A) A one-dimensional monitored quantum circuit is mapped to a directed spacetime graph $G$, where nodes correspond to spacetime events $(i,t)$ and edges encode the causal structure of the circuit through worldline and gate-induced connections. Each node carries features derived from the measurement record, including a measurement indicator, measurement outcome, normalized time coordinate, and positional information. (B) Single-scale graph neural network architecture. Stacking $K$ directed GraphSAGE message-passing layers enlarges the causal receptive field linearly with depth, corresponding to a $K$-layer causal light cone on the spacetime graph. (C) RG-inspired hierarchical architecture. Iterative $2\times2$ spacetime blocking generates progressively coarser graphs, enabling the network to access larger effective spacetime regions and increasing the receptive field exponentially with the number of coarse-graining levels. (D) Regression task. Node embeddings at the final time slice are pooled separately over the left and right halves of the chain to obtain global representations, which are combined by a multilayer perceptron (MLP) to predict the normalized half-chain von Neumann entropy $s=S_{N/2}/(N/2)$. (E) Prediction and training objective. The model prediction $\hat{s}$ is trained to match the target entropy using the mean squared error loss $\mathcal{L}=(\hat{s}-s)^2$.
}

\label{fig:fig1}
\end{figure*}

\section{Model and methods}
\label{sec:II}

\subsection{Monitored random quantum circuits}
\label{susec:mrqc}

We study monitored one-dimensional random quantum circuits with open boundary conditions. Each circuit consists of $N$ qubits and has depth $t_{\max}=30$,
which is sufficient for the entanglement dynamics to approach a
steady scaling regime for the system sizes considered. At each time layer, two-qubit Haar-random gates are applied in a brick-wall pattern: even layers act on bonds $(0,1),(2,3),\dots$, while odd layers act on $(1,2),(3,4),\dots$. After each unitary layer, each qubit is independently projectively measured in the computational basis with probability $p$. A schematic illustration of the monitored circuit is shown in Fig.~\ref{fig:fig1}.

For system sizes $N \in \{8,10,12,14, 16\}$, the circuit dynamics are simulated using exact state-vector evolution. For larger systems, matrix product state simulations are performed using ITensor~\cite{itensor} with adaptive bond dimension to control truncation error. The circuit is initialized in the product state $|0\cdots0\rangle$.

For each circuit realization, the supervised learning target is the half-chain von Neumann entropy at the final time $t_{\max}$,
\begin{equation}
S_{N/2}(t_{\max}) = -\mathrm{Tr}\,\rho_L \log \rho_L ,
\end{equation}
where $\rho_L$ denotes the reduced density matrix of the left half of the chain.

To improve generalization across system sizes, we train on the normalized entropy
\begin{equation}
s = \frac{S_{N/2}}{N/2}.
\end{equation}
This normalization removes the leading extensive scaling with system size and allows the learning task to focus on the underlying intensive structure of the entanglement.

\subsection{Spacetime graph representation}
\label{subsec:sgr}

Each circuit realization is mapped to a directed spacetime graph $G=(V,E)$ following the message-passing framework~\cite{gilmer2017,hamilton2018,kipf2017}. Nodes correspond to spacetime events $(i,t)$ with $i\in\{0,\dots,N-1\}$ and $t\in\{0,\dots,t_{\max}\}$. Directed edges encode causal propagation from past to future. Worldline edges connect $(i,t)\rightarrow(i,t+1)$. For each two-qubit gate acting on bond $(i,i+1)$ at time $t$, diagonal edges $(i,t)\rightarrow(i+1,t+1)$ and $(i+1,t)\rightarrow(i,t+1)$ are included. No future-to-past edges are introduced. This construction ensures that message passing respects the causal structure of the quantum circuit.

Each node $(i,t)$ carries a feature vector
\begin{equation}
x_{(i,t)} = (m_{i,t},\, y_{i,t},\, t/t_{\max},\, x_L,\, x_R),
\end{equation}
where $m_{i,t}\in\{0,1\}$ is a measurement indicator, $y_{i,t}\in\{-1,+1\}$ is the measurement outcome (set to zero if no measurement occurs). Note that the measurement indicator $m_{i,t}$ ensures that the outcome $y_{i,t}=0$ is unambiguously interpreted as “no measurement”. 
and
\begin{equation}
x_L=\frac{i}{N-1}, \qquad x_R=1-\frac{i}{N-1}
\end{equation}
encode boundary-aware positional information. Raw qubit indices are not included as features, enabling transferability across system sizes under open boundary conditions. The normalized time coordinate provides the model with global information about circuit depth

\subsection{Single-scale directed graph neural network}
\label{subsec:single_scale}

We first consider a Single-scale directed GraphSAGE architecture~\cite{hamilton2018}. For node $v$, the $\ell$-th layer update is
\begin{equation}
h_v^{(\ell+1)} =
\phi\!\left(
W_0 h_v^{(\ell)} +
\frac{1}{|\mathcal{N}^-(v)|}
\sum_{u\in\mathcal{N}^-(v)} W_1 h_u^{(\ell)}
\right),
\end{equation}
where $\mathcal{N}^-(v)$ denotes incoming neighbors and $\phi$ is a ReLU activation. Layer normalization is applied after each layer. The hidden dimension is fixed to $128$. The total depth of the Single-scale architecture is denoted by $K$ and varied in $\{1,2,3,4,6\}$.

Readout is performed using node embeddings at the final time slice $t=t_{\max}$. Let $\mathcal{L}=\{i<N/2\}$ and $\mathcal{R}=\{i\ge N/2\}$ denote the left and right halves. Mean pooling yields
\begin{equation}
z_L = \frac{1}{|\mathcal{L}|} \sum_{i\in\mathcal{L}} h_{(i,t_{\max})},
\qquad
z_R = \frac{1}{|\mathcal{R}|} \sum_{i\in\mathcal{R}} h_{(i,t_{\max})}.
\end{equation}
The entropy prediction is obtained via a multilayer perceptron acting on
\begin{equation}
\hat{s} = \mathrm{MLP}\!\left([z_L,\, z_R,\, |z_L-z_R|,\, z_L\odot z_R]\right).
\end{equation}
This combination captures differences and products between halves, which are natural features for predicting entanglement.

\subsection{Single-scale locality and finite-scale interpretation}

The depth $K$ of the single-scale network admits a natural interpretation in terms of locality and causal propagation. In strictly unitary local quantum systems, Lieb–Robinson bounds~\cite{Lieb1972,nachtergaele2010} imply
\begin{equation}
\|[O_x(t),O_y(0)]\|\le 
C \exp\!\left(-\frac{|x-y|-v_{\mathrm{LR}}t}{\zeta}\right),
\end{equation}
which expresses exponential suppression of correlations outside the light cone $|x-y|\le v_{\mathrm{LR}}t$.

In monitored circuits, the presence of projective measurements breaks unitarity, so rigorous Lieb–Robinson bounds derived for purely unitary dynamics do not directly apply. Nevertheless, the local two-qubit gates impose a finite causal propagation speed determined by the circuit geometry. In the brick-wall circuit used here, information can spread by at most one lattice site per circuit layer. This defines an effective lattice light cone with velocity $v_{\mathrm{lat}} = 1$, in lattice units.

A depth-$K$ directed GNN aggregates information only from nodes connected by directed paths of length at most $K$. The corresponding truncated backward light cone of a node $(i_*,t_*)$ is therefore contained within
\begin{equation}
\mathcal{R}_K(i_*,t_*) \subseteq 
\left\{(i,t): t_*-K\le t\le t_*,\;
|i-i_*|\le (t_*-t)\right\}.
\end{equation}
Thus, the effective spatial scale resolved by the single-scale architecture grows approximately linearly with depth,
\begin{equation}
\ell_{\mathrm{eff}}^{\mathrm{single}}(K)\sim K .
\end{equation}

Near the measurement-induced transition, the circuit exhibits a correlation length
\begin{equation}
\xi(p)\sim |p-p_c|^{-\nu}.
\end{equation}
Since the single-scale network accesses correlations only up to $\ell_{\mathrm{eff}}^{\mathrm{single}}(K)$, we hypothesize that its prediction error may depend on the ratio $\ell_{\mathrm{eff}}^{\mathrm{single}}(K)/\xi(p)$. This motivates a phenomenological finite-scale form
\begin{equation}
\varepsilon(K,p)=K^{-\alpha}
f\!\left(\frac{\ell_{\mathrm{eff}}^{\mathrm{single}}(K)}{\xi(p)}\right),
\end{equation}
where $\alpha$ and $f$ are determined empirically. This scaling form is treated as a hypothesis and tested against numerical data in Sec.~\ref{sec:III}. In this operational sense, the network depth $K$ acts as an infrared cutoff, limiting the spacetime scale accessible to the model.

\subsection{RG-inspired hierarchical graph neural network}
\label{subsec:rg_hgnn}

To make the coarse-graining interpretation explicit at the architectural level, we introduce a hierarchical spacetime GNN based on iterative $2\times2$ blocking.

Starting from the original graph $G^{(0)}$, we define the blocking transformation
\begin{equation}
(i,t)\longmapsto(i',t'), \qquad
i'=\left\lfloor\frac{i}{2}\right\rfloor,\quad
t'=\left\lfloor\frac{t}{2}\right\rfloor.
\end{equation}
Each coarse node aggregates the feature vectors of the four fine nodes in its spacetime block via mean pooling. Directed edges between coarse nodes are induced whenever at least one fine-scale edge connects the corresponding blocks. Duplicate and self-loop edges are removed, producing a new coarse graph $G^{(1)}$. Repeated application generates a hierarchy
\begin{equation}
G^{(0)}\rightarrow G^{(1)}\rightarrow G^{(2)}\rightarrow\cdots.
\end{equation}

On each level $\ell$, we apply $K_\ell$ directed GraphSAGE layers before proceeding to the next coarse level. The hierarchical architecture is parameterized by
\begin{equation}
\mathbf{K}=(K_0,K_1,\dots,K_{L-1}),
\end{equation}
where $L$ is the number of hierarchical levels. The RG depth is defined as $J=L-1$, equal to the number of coarse-graining steps. In our implementations we consider $\mathbf{K}=(2,2)$, corresponding to $J=1$, and $\mathbf{K}=(2,2,2)$, corresponding to $J=2$.

After $J$ blocking steps, the effective spacetime resolution scales approximately as
\begin{equation}
N\rightarrow \frac{N}{2^J},\qquad
t_{\max}\rightarrow \frac{t_{\max}}{2^J}.
\end{equation}
Message passing at the coarsest level probes correlations between blocks of linear size $2^J$, yielding an effective scale
\begin{equation}
\ell_{\mathrm{eff}}^{\mathrm{RG}}(J)\sim 2^J K_{\mathrm{top}},
\end{equation}
where $K_{\mathrm{top}}$ denotes the number of message-passing layers applied at the coarsest level.

Thus, unlike the single-scale architecture where the accessible scale grows linearly with depth, the hierarchical model increases the effective infrared scale exponentially with the RG depth $J$. While this construction does not constitute an exact renormalization group transformation in the field-theoretic sense, it provides an operational real-space coarse-graining of the spacetime graph. This enables a direct comparison between enlarged receptive field (single-scale model) and explicit blocking (hierarchical model) in probing correlations near the measurement-induced transition.

We emphasize that the blocking transformation operates on the spacetime graph representation and does not correspond to a coarse-graining of the underlying quantum state or density matrix. The hierarchical architecture should therefore be interpreted as an architectural coarse-graining of measurement records rather than a microscopic RG transformation. Its relevance to RG ideas lies in the controlled modification of the effective spacetime scale accessible to the model.

\subsection{Training protocol}

Training is performed on mixed system sizes $N\in\{8,10,12,14,16\}$ using an
80/20 train/validation split at the circuit-realization level. The loss
function is the mean squared error
\begin{equation}
\mathcal{L} = \frac{1}{B}\sum_{i=1}^B (\hat{s}_i - s_i)^2,
\end{equation}
optimized using the Adam optimizer with learning rate $10^{-3}$ and
weight decay $10^{-5}$. Mini-batches contain graphs of varying sizes;
batching is handled using PyTorch Geometric's dynamic batching
mechanism, which preserves graph boundaries.

During training the model checkpoint with the lowest validation loss
is retained. After training, the model is evaluated without fine-tuning
on larger systems generated using matrix product state (MPS)
simulations.

For comparison between single-scale and hierarchical models,
architectures are chosen such that the total number of message-passing
layers is comparable (e.g., single-scale $K=6$ versus hierarchical
$(2,2,2)$). For hidden dimension 128, the single-scale $K=6$ model
contains approximately $2\times10^5$ trainable parameters, comparable
to the hierarchical $(2,2,2)$ model, which has the same number of
message-passing layers and therefore nearly identical parameter count.
The blocking operations in the hierarchical architecture introduce no
additional trainable parameters.

\begin{figure}[t]
    \centering
    \includegraphics[width=0.95\columnwidth]{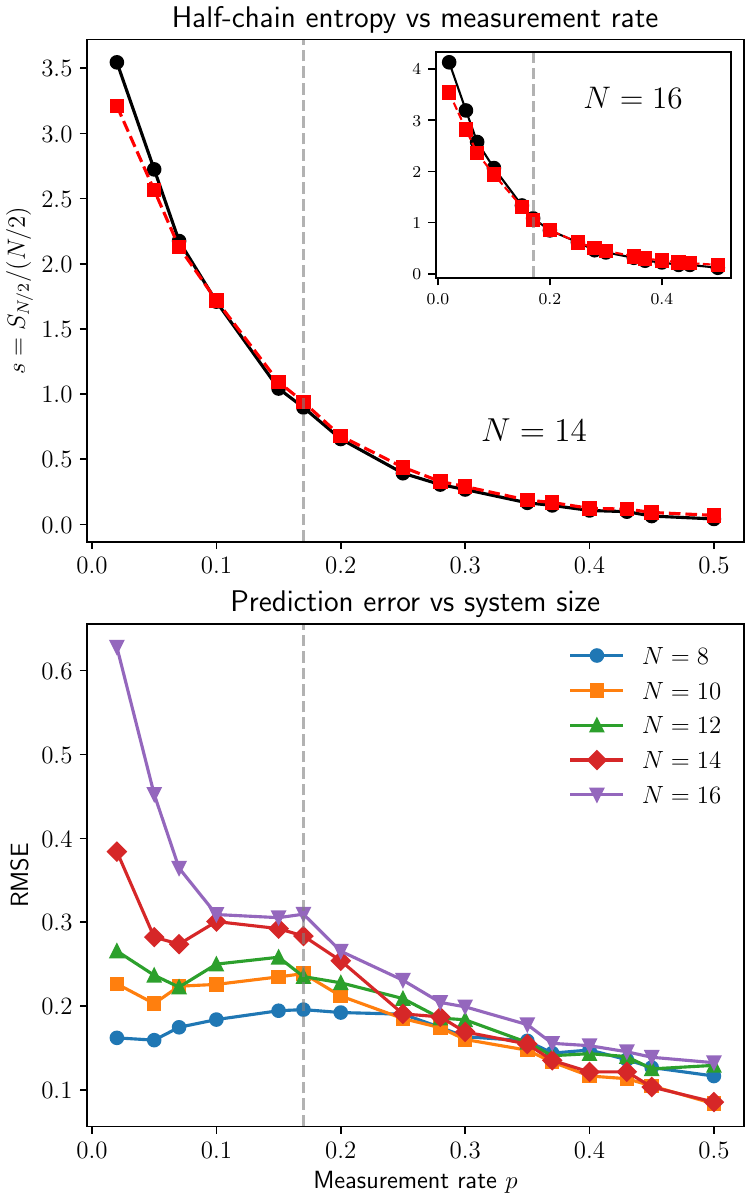}
  \caption{
Baseline behavior of the monitored circuit and prediction accuracy of the graph neural network.
Results are obtained using the single-scale GNN with depth $K=6$.
(\textbf{Top}) Normalized half-chain entropy $s = S_{N/2}/(N/2)$ as a function of measurement rate $p$ for system sizes $N=14$ (main panel) and $N=16$ (inset). Black circles show the exact values obtained from quantum circuit simulations, while red squares denote neural-network predictions. The dashed vertical line indicates the approximate location of the measurement-induced transition at $p_c \approx 0.17$.
(\textbf{Bottom}) Root-mean-square prediction error (RMSE) as a function of measurement rate for several system sizes. The prediction error is largest in the weak-measurement regime where entanglement is extensive and long-range correlations dominate, and decreases significantly at larger $p$ where frequent measurements suppress entanglement growth.
}
\label{fig:fig1}
\end{figure}

\section{Results}
\label{sec:III}

In this section we analyze how accurately graph neural networks can
reconstruct global entanglement observables from local spacetime
measurement records in monitored quantum circuits. Our primary goal is
to determine how the prediction accuracy depends on the spacetime scale
accessible to the learning architecture.

We proceed in several steps. First, we verify that the neural network
accurately reproduces the dependence of half-chain entanglement entropy
on the measurement rate. We then investigate how the prediction error
depends on the depth of the single-scale GNN, which controls the size
of the spacetime region over which information can be aggregated.
Finally, we compare this single-scale architecture with the
hierarchical model introduced in Sec.~\ref{subsec:rg_hgnn} and examine
how prediction accuracy scales with the effective spacetime scale
accessible to the network.

\subsection{Entanglement structure and baseline prediction accuracy}
\label{subsec:baseline}

Figure~\ref{fig:fig1}  shows results obtained using the single-scale architecture with depth $K=6$. The upper panel
shows the normalized half-chain entropy $s = S_{N/2}/(N/2)$ as a function of the measurement rate $p$ for system sizes
$N=14$ and $N=16$. While $N=14$ is included in the training set,
the system size $N=16$ is not used during training and therefore
provides a direct test of the model's ability to generalize to
larger systems. As expected for monitored random circuits,
the entanglement entropy decreases continuously as the measurement
rate increases. At small $p$, the system resides in a volume-law
entangled phase characterized by large bipartite entanglement,
whereas at large $p$ the entropy approaches an area-law regime due
to frequent measurements that suppress entanglement growth.
The dashed vertical line indicates the approximate location of the
measurement-induced transition at $p_c \approx 0.17$.

The neural network predictions (red symbols) closely track the exact
entanglement values obtained from quantum simulations (black symbols)
across the full range of measurement rates. Importantly, this agreement
persists even for the unseen system size $N=16$, demonstrating that the
model generalizes beyond the system sizes used during training. This
result indicates that the network successfully reconstructs global
entanglement observables using only local spacetime measurement data
as input, without access to the underlying quantum state. The ability
to generalize to larger system sizes suggests that the relevant
information for predicting entanglement is encoded in local spacetime
patterns of measurements rather than in system-size–specific features.

The lower panel of Fig.~\ref{fig:fig1} shows the root-mean-square
prediction error as a function of measurement rate for several system
sizes. The error exhibits a clear dependence on both measurement rate
and system size. In particular, the prediction error is largest in the
weak-measurement regime, where entanglement spreads over large
spacetime regions, and decreases substantially as the measurement rate
increases. This behavior is consistent with the physical intuition
that reconstructing volume-law entanglement requires integrating
information over larger spacetime scales than in the strongly
measured phase.

We also observe that the prediction error increases with system size
for fixed measurement rate, reflecting the growing complexity of the
entanglement structure in larger systems. These observations motivate
a more detailed investigation of how the accessible spacetime scale
of the learning architecture influences prediction accuracy.

\begin{figure}[t]
    \centering
    \includegraphics[width=\columnwidth]{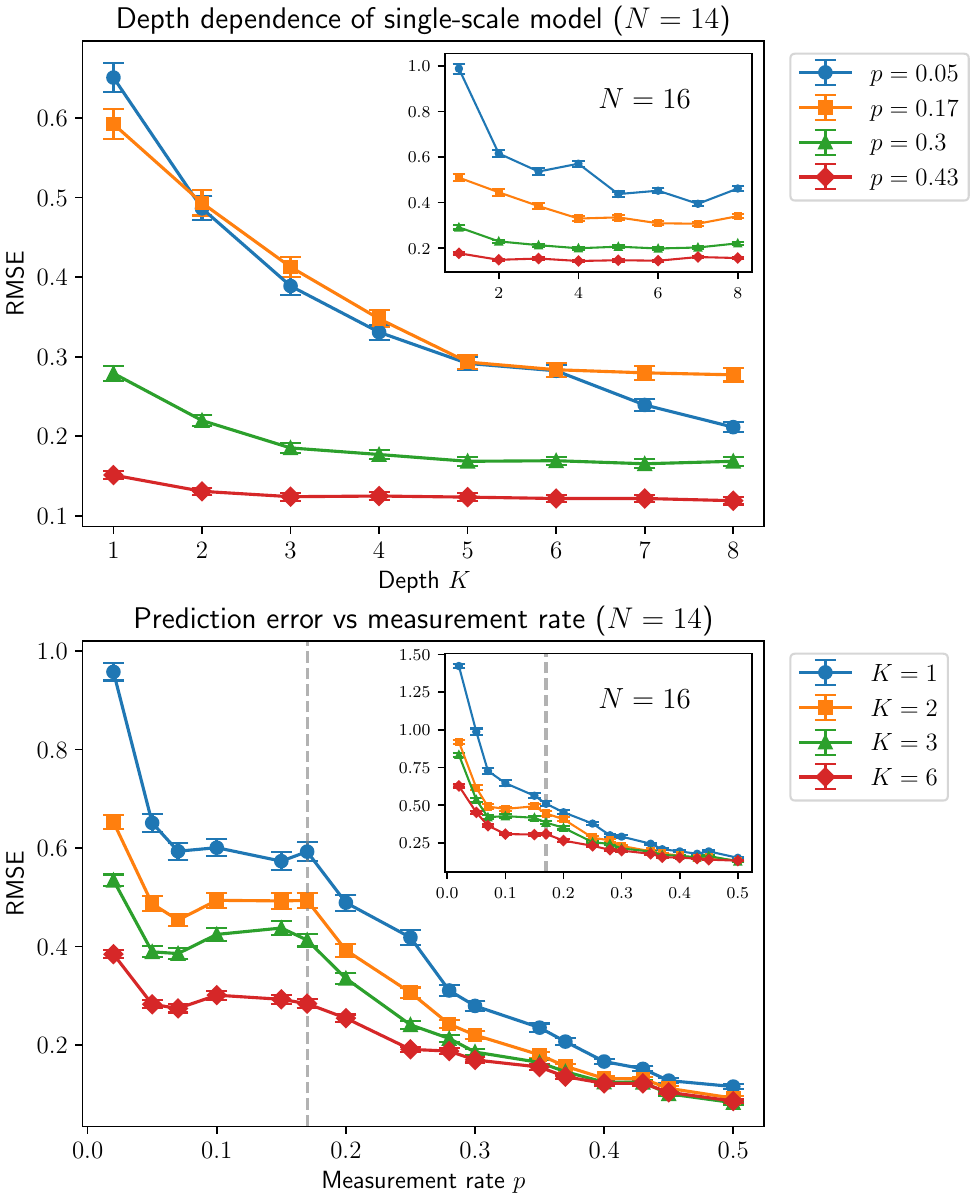}
\caption{
Depth dependence of the single-scale graph neural network. (\textbf{Top}) Root-mean-square prediction error (RMSE) as a function of network depth $K$ for several measurement rates, shown for system size $N=14$. The inset shows the corresponding results for the unseen system size $N=16$, demonstrating that the depth dependence generalizes beyond the system sizes used during training. (\textbf{Bottom}) Prediction error as a function of measurement rate for several network depths at $N=14$. The dashed vertical line indicates the approximate measurement-induced transition $p_c\approx0.17$. The prediction error is largest in the weak-measurement regime and decreases with increasing depth, reflecting the need to integrate information over larger spacetime scales when entanglement is extensive. Insets again show the corresponding behavior for the unseen system size $N=16$.
}
\label{fig:fig2}
\end{figure}

\begin{figure*}[t]
    \centering
    \includegraphics[width=\linewidth]{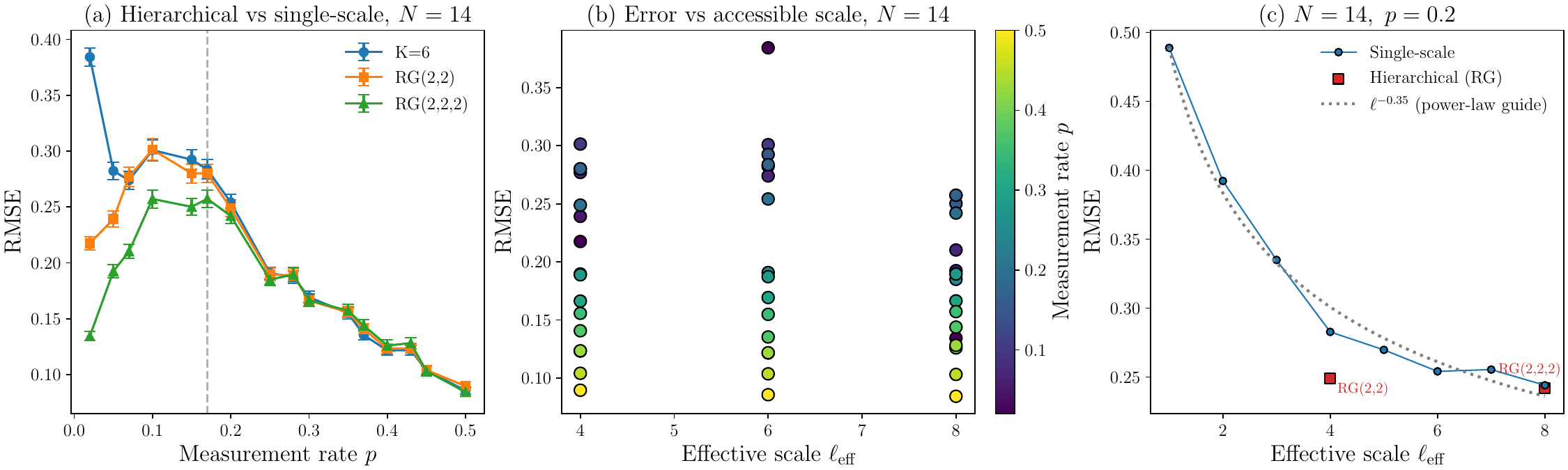}
\caption{Comparison between single-scale and hierarchical graph neural
network architectures and their dependence on the accessible spacetime scale. (\textbf{a}) Prediction error as a function of measurement rate
for the single-scale architecture with depth $K=6$ and for two
hierarchical architectures RG$(2,2)$ and RG$(2,2,2)$ at system size
$N=14$. All models have comparable numbers of message-passing layers
but access different effective spacetime scales. The dashed vertical
line indicates the approximate measurement-induced transition
$p_c\approx0.17$. (\textbf{b}) Prediction error plotted as a function of the effective
spacetime scale $\ell_{\mathrm{eff}}$ defined in
Sec.~\ref{subsec:rg_hgnn}. Each point corresponds to a different
architecture and measurement rate. Colors indicate the measurement
rate $p$. The error decreases systematically with increasing
accessible scale, largely independent of the architectural details. (\textbf{c}) Scaling of the prediction error with accessible scale
for a representative measurement rate $p=0.2$. Blue circles show the
single-scale architecture for depths $K=1\ldots8$, while red squares
show the hierarchical architectures. The dashed curve indicates a
power-law guide $\varepsilon \sim \ell_{\mathrm{eff}}^{-0.35}$.}
\label{fig:fig3}
\end{figure*}

\begin{figure*}[t]
    \centering
    \includegraphics[width=\linewidth]{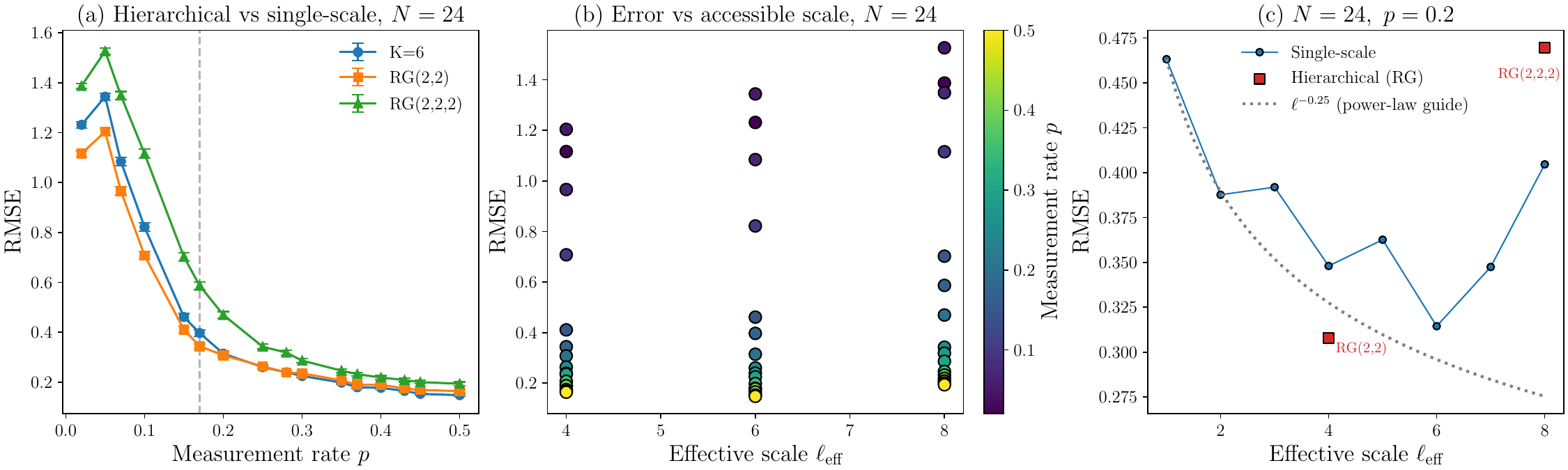}
\caption{Same as Fig~\ref{fig:fig3}, scaling of prediction error with accessible spacetime scale for an
unseen system size $N=24$. Exact data were generated using
tensor-network simulations. The network was trained only on
systems with $N\leq14$.}
\label{fig:fig3b}
\end{figure*}

\subsection{Depth dependence of the single-scale architecture}
\label{subsec:depth}

We next investigate how the prediction accuracy depends on the depth
of the single-scale graph neural network. As discussed in
Sec.~\ref{subsec:single_scale}, the network depth $K$ determines the
maximum number of message-passing steps and therefore controls the
size of the spacetime neighborhood from which information can be
aggregated. Increasing the depth effectively enlarges the causal
region accessible to the model, providing an operational probe of the
spacetime scale required to reconstruct global entanglement.

The upper panel of Fig.~\ref{fig:fig2} shows the root-mean-square
prediction error (RMSE) as a function of network depth for several
measurement rates at system size $N=14$. The prediction error decreases
systematically with increasing depth, indicating that deeper networks
are able to integrate information over larger spacetime regions.
This effect is particularly pronounced in the weak-measurement regime
($p=0.05$), where the system exhibits strong volume-law entanglement.
In this regime, shallow networks that access only small spacetime
neighborhoods incur large prediction errors, whereas increasing the
depth significantly improves the accuracy.

By contrast, in the strongly measured regime ($p=0.43$) the prediction
error is already small for shallow networks and shows only weak
dependence on depth. This behavior reflects the reduced spatial
extent of entanglement when frequent measurements suppress the growth
of quantum correlations. In this regime, the information required to
predict the half-chain entropy is largely contained within relatively
local spacetime regions.

The inset of the upper panel shows the corresponding depth dependence
for the unseen system size $N=16$. Despite the model being trained only
on system sizes $N\le14$, the qualitative behavior remains essentially
unchanged. The systematic reduction of prediction error with increasing
depth therefore generalizes beyond the system sizes used during
training, indicating that the relation between network depth and
accessible spacetime scale captures intrinsic properties of the
monitored circuit dynamics rather than system-size–specific features.

The lower panel of Fig.~\ref{fig:fig2} further illustrates this behavior
by plotting the prediction error as a function of measurement rate for
several fixed depths. For shallow networks ($K=1$ or $K=2$), the
prediction error is large in the weak-measurement regime and decreases
gradually as the measurement rate increases. As the depth grows, the
error decreases across the entire range of measurement rates, with the
largest improvements occurring near and below the measurement-induced
transition at $p_c\approx0.17$. Insets again show the corresponding
results for $N=16$, confirming that the same qualitative trends persist
for unseen system sizes.

Taken together, these results support the interpretation that the
prediction accuracy of the single-scale architecture is controlled
primarily by the spacetime scale accessible to the model. Increasing
the network depth enlarges this scale approximately linearly,
consistent with the relation $\ell_{\mathrm{eff}}^{\mathrm{single}} \sim K$ 
introduced in Sec.~\ref{subsec:single_scale}. This observation
motivates exploring architectures that can access larger spacetime
scales more efficiently, which we address in the following section
using a hierarchical coarse-graining approach.

\subsection{Hierarchical architectures and scaling with accessible spacetime scale}

We now compare the single-scale architecture with the hierarchical
graph neural network introduced in
Sec.~\ref{subsec:rg_hgnn}, which combines local message passing with
explicit spacetime blocking. The hierarchical model allows the
network to access larger effective spacetime regions without
requiring a proportionally larger number of message-passing layers.

Figure~\ref{fig:fig3}(a) shows the prediction error as a function of
measurement rate for the single-scale architecture with depth
$K=6$ together with two hierarchical models, RG$(2,2)$ and
RG$(2,2,2)$. Across a broad range of measurement rates, the
hierarchical architectures achieve prediction accuracy comparable to
that of the single-scale model. This is notable because the
hierarchical architectures access larger effective spacetime regions
through coarse-graining rather than by increasing the number of
message-passing steps.

To explore whether prediction accuracy is related to the accessible
spacetime scale, we replot the errors as a function of the effective
scale $\ell_{\mathrm{eff}}$ defined in
Sec.~\ref{subsec:rg_hgnn}. As shown in Fig.~\ref{fig:fig3}(b), the
prediction error generally decreases as $\ell_{\mathrm{eff}}$
increases. In the present comparison the architectures correspond to
three discrete effective scales,
$\ell_{\mathrm{eff}}=4,6,8$, and points at each scale represent
results obtained for different measurement rates. Although the
scatter reflects the dependence on measurement rate, models with
larger accessible spacetime regions tend to achieve lower prediction
errors irrespective of the architectural details.

Figure~\ref{fig:fig3}(c) further illustrates this trend for a
representative measurement rate $p=0.2$. The prediction error
decreases with increasing effective scale and is roughly consistent
with a power-law trend $\varepsilon \sim
\ell_{\mathrm{eff}}^{-\alpha}$ over the range of scales explored
here, with $\alpha \approx 0.35$ for the present dataset. The
hierarchical architectures fall close to the same curve obtained for
the single-scale models, suggesting that the prediction accuracy is
primarily influenced by the spacetime scale accessible to the
network rather than by the specific architectural implementation.

Taken together, these results are consistent with the view that
reconstructing global entanglement observables requires integrating
measurement information over progressively larger spacetime regions.
Architectures that access larger effective scales correspondingly
tend to achieve improved predictive performance.

To test whether the observed relationship between prediction
accuracy and accessible spacetime scale persists beyond the
system sizes used during training, we repeat the analysis for
a larger system with $N=24$. This size was not included in the
training data and therefore provides a stringent test of the
model's ability to generalize.

Figure~\ref{fig:fig3b} presents the same analysis for system size
$N=24$, where the exact data were generated using tensor-network
simulations. Because this system size was not included during
training, the prediction errors are generally larger than those
observed for $N=14$. Nevertheless, several qualitative features
remain similar.

First, the hierarchical architectures achieve prediction accuracy
comparable to that of the single-scale network across a broad range
of measurement rates [Fig.~\ref{fig:fig3b}(a)]. Second, when the
prediction error is plotted as a function of the effective
spacetime scale [Fig.~\ref{fig:fig3b}(b)], models with larger
accessible scales again tend to exhibit smaller errors despite the
increased scatter associated with the larger system size.

Finally, Fig.~\ref{fig:fig3b}(c) shows the dependence of the error on
$\ell_{\mathrm{eff}}$ for a representative measurement rate
$p=0.2$. Although the scaling is less regular than in the $N=14$
case, the hierarchical architectures remain broadly consistent with
the trend observed for the single-scale models. These results
suggest that the relationship between prediction accuracy and the
accessible spacetime scale persists even for system sizes that lie
outside the training distribution.

\begin{figure*}[t]
    \centering
    \includegraphics[width=\linewidth]{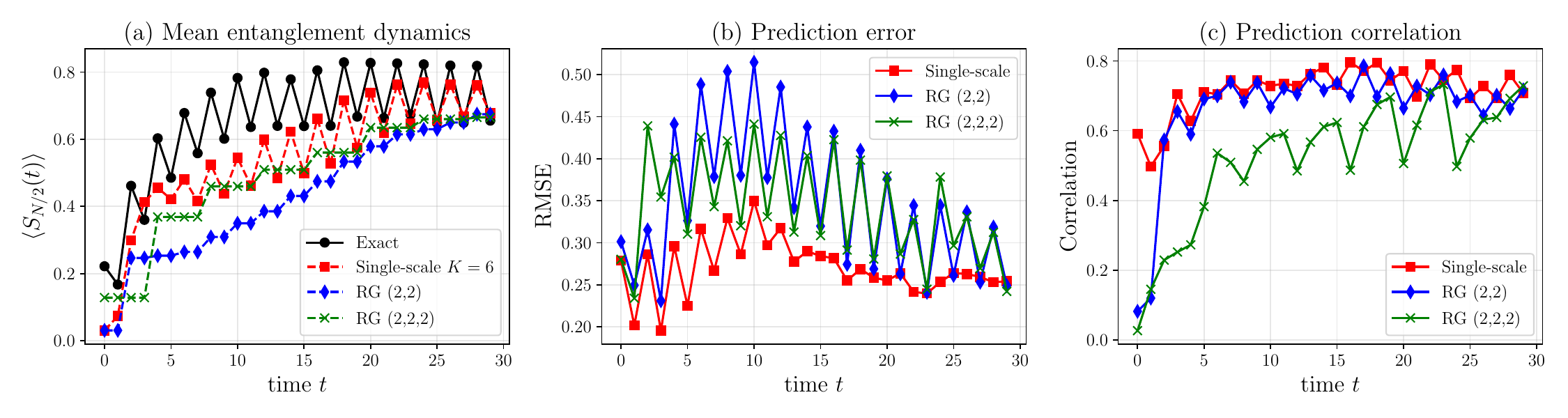}
\caption{Off-target-time readout of entanglement dynamics for system size
$N=14$ and measurement rate $p=0.4$. The models are trained only on
the final-time entropy at $t_{\max}=30$, but are evaluated here by
applying the learned readout to intermediate time slices. (\textbf{a}) Mean half-chain entropy
$\langle S_{N/2}(t)\rangle$ as a function of time for the exact
dynamics, the single-scale model with depth $K=6$, and the two
hierarchical architectures RG$(2,2)$ and RG$(2,2,2)$. All models
capture the overall entanglement growth and saturation, with the
single-scale model showing the closest agreement to the exact
trajectory. (\textbf{b}) Time-resolved root-mean-square prediction error.
The hierarchical models exhibit larger deviations at short and
intermediate times, consistent with their reduced temporal resolution
due to coarse-graining. (\textbf{c}) Pearson correlation between predicted and exact entropies across circuit realizations at each time. Despite being
trained only on the final-time target, all models retain substantial
trajectory-level correlation over much of the evolution.}
\label{fig:fig4}
\end{figure*}

\subsection{Off-target-time readout of entanglement dynamics}
\label{subsec:dynamics}

Although the models considered here are trained only to predict the
half-chain entropy at the final time $t_{\max}$, the learned spacetime
representations can also be probed at earlier times by applying the
same readout procedure to intermediate time slices. This provides a
useful test of whether the latent representation learned under
final-time supervision retains information about the temporal buildup
of entanglement.

Figure~\ref{fig:fig4} shows the resulting off-target-time predictions
for a representative case in the strongly measured regime,
$N=14$ and $p=0.4$. Panel~(a) compares the exact mean entanglement
trajectory $\langle S_{N/2}(t)\rangle$ with predictions obtained from
the single-scale model and from the two hierarchical architectures.
All three models reproduce the overall growth and late-time saturation
of the entanglement entropy, despite having been trained only on the
final-time target. The single-scale architecture provides the most
accurate reconstruction of the full trajectory, while the hierarchical
models capture the coarse trend but smooth over part of the short-time
temporal structure.

This difference is reflected in the time-resolved error shown in
Fig.~\ref{fig:fig4}(b). The single-scale model exhibits the smallest
root-mean-square error over most of the evolution, whereas the
hierarchical models show larger deviations at early and intermediate
times. These deviations are consistent with the explicit temporal
coarse-graining built into the hierarchical architectures, which
reduces sensitivity to short-time oscillatory features while preserving
larger-scale structure.

Panel~(c) shows the Pearson correlation between predicted and exact
entropies across individual circuit realizations at each time. The
correlation rapidly increases during the evolution and remains
substantial over most of the trajectory for all architectures,
reaching values of order $0.7$--$0.8$ for the single-scale and
RG$(2,2)$ models at intermediate and late times. Even the deeper
hierarchical model RG$(2,2,2)$, while less accurate at short times,
retains significant predictive correlation once the entanglement
growth has developed.

Taken together, these results indicate that final-time supervision
induces a latent spacetime representation that contains information
about the broader entanglement dynamics, not only the terminal
observable used for training. At the same time, the reduced
performance of the hierarchical models at short times highlights that
off-target-time prediction is sensitive to the temporal resolution of
the architecture and should therefore be interpreted as a coarse
diagnostic rather than as a fully supervised reconstruction of the
entire trajectory.

\section{Summary}
\label{sec:summary}

In this work we investigated whether global entanglement properties
of monitored quantum circuits can be reconstructed directly from
local measurement records using graph neural networks.
The spacetime structure of the measurement data naturally defines
a graph representation, allowing message-passing architectures to
aggregate information across both space and time.

We first demonstrated that a graph neural network can accurately
predict the half-chain entanglement entropy of monitored random
circuits over a broad range of measurement rates.
The model reproduces the expected dependence of entanglement on
the measurement rate and remains accurate across different system
sizes. Importantly, the network generalizes to larger system sizes
that were not included during training, indicating that the learned
representation captures system-size–independent spacetime features
of the measurement dynamics.

We then examined how the architecture controls the spacetime scale
accessible to the model. Increasing the depth of a single-scale
network reduces the prediction error, consistent with the picture
that deeper message passing allows the model to integrate
information from larger spacetime regions.
Introducing hierarchical coarse-graining further extends the
effective spacetime scale that can be accessed with a limited
number of layers. The prediction error collapses when plotted
against this effective scale, suggesting that the dominant factor
governing prediction accuracy is the spacetime region from which
the model can collect information.

Finally, we showed that the learned representations retain
information about the temporal buildup of entanglement.
Although the networks were trained only on the final-time entropy,
the same readout procedure applied to intermediate time slices
recovers the overall growth of the entanglement entropy and
maintains significant correlation with the exact dynamics.
Influence maps further reveal that the networks rely on localized
spacetime regions whose extent grows with architectural depth,
providing a qualitative visualization of the effective spacetime
scale explored by the model.

These results highlight the potential of graph-based learning
approaches for extracting physically meaningful observables from
local measurement data in quantum many-body systems.
More broadly, they suggest that machine-learning architectures can
serve as controlled probes of the spacetime structure underlying
complex quantum dynamics.

Several directions for future work remain open. It would be
interesting to apply similar approaches to other observables,
including operator spreading and correlation functions, and to
investigate whether such models can identify critical behavior
near measurement-induced phase transitions. Recent work has also
highlighted the rich structure of measurement-induced information,
including finite-time teleportation transitions in monitored
circuits \cite{Bao2024} and universal long-distance
features of measurement-induced entanglement in many-body quantum
states \cite{Cheng2024}. Extending the present reconstruction
framework to probe such phenomena directly from measurement
records could provide new insights into the spacetime organization
of measurement-induced correlations. Another promising direction is
to explore adaptive or physics-informed hierarchical architectures
that more closely mirror the underlying causal structure of
monitored quantum circuits \cite{Ivaki2025}.

\section{Acknowledgments}
J.V. acknowledges the Ataro Group for generously providing office space and computational resources that facilitated the completion of this work.

\appendix
\section{Measurement probability under spacetime blocking}
\label{app:measurement_rg}

In the hierarchical architecture introduced in Sec.~\ref{subsec:rg_hgnn}, the spacetime
graph representation of a monitored circuit is coarse-grained through
iterative $2\times2$ blocking. This construction operates on the graph
representation used by the neural network and does not define a
renormalization-group transformation of the underlying circuit
dynamics. In particular, the measurement probability $p$ of the
monitored circuit is not rescaled under this architectural
coarse-graining.

It is nevertheless instructive to contrast this construction with a
hypothetical renormalization of the circuit itself. If two consecutive
time layers of the circuit were coarse-grained into a single effective
layer, each qubit would experience two independent opportunities for a
projective measurement. In that case the probability of observing at
least one measurement event would be
\begin{equation}
p' = 1-(1-p)^2,
\end{equation}
which corresponds to the probability that at least one of the two
measurement attempts occurs. For small $p$, this yields $p' \approx 2p$.
Such a transformation would therefore modify the effective measurement
rate of the circuit.

The hierarchical graph construction used in this work differs
fundamentally from this scenario. The coarse-graining step groups
multiple spacetime nodes of the measurement record into a single
coarse node, but it does not generate a new circuit ensemble or
modify the physical measurement process. Instead, the feature vector
associated with a coarse node aggregates the features of the
corresponding fine nodes (via mean pooling in our implementation).
Consequently, the network receives an averaged summary of the
measurement information within each spacetime block rather than a
renormalized measurement probability.

In this sense the blocking procedure should be interpreted as an
architectural coarse-graining of the spacetime graph representation
rather than a renormalization of the monitored circuit itself.
The physical parameter $p$ therefore remains unchanged throughout the
hierarchical construction, while the effective spacetime scale
accessible to the model increases with the RG depth.

\begin{figure*}[t]
\centering
\includegraphics[width=\linewidth]{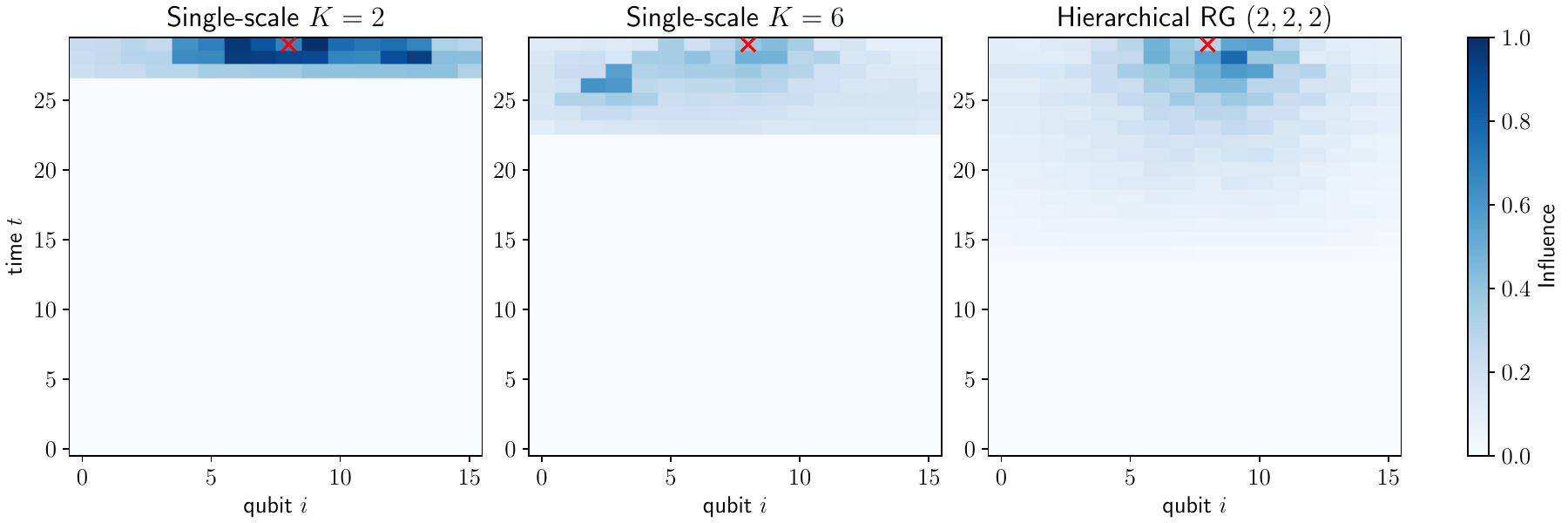}\\
\includegraphics[width=\linewidth]{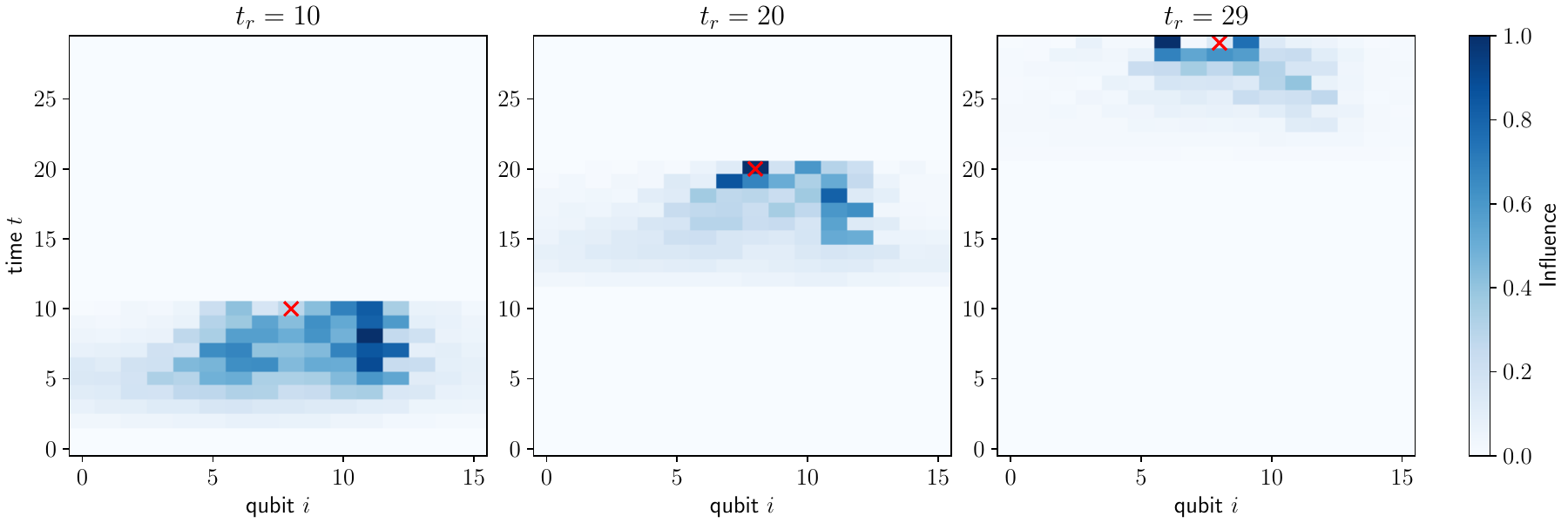}
\caption{
Spacetime influence maps illustrating which measurement events
$(i,t)$ contribute most strongly to the neural-network prediction of
the half-chain entropy.
(\textbf{Top}) Influence maps for three representative architectures at $t_r=30$:
a shallow single-scale model ($K=2$), a deeper single-scale model
($K=6$), and a hierarchical RG$(2,2,2)$ model. The red cross marks the
readout location near the entanglement cut.
(\textbf{Bottom}) Time-resolved influence maps obtained by evaluating
the sensitivity of the prediction at different readout times
$t_r=10,20,29$ using single-scale model ($K=6$). The dominant influence region shifts upward in time
and expands as the readout time increases, indicating that the model
integrates information from a growing spacetime region of the
measurement record.
}
\label{fig:influence_maps}
\end{figure*}

\section{Spacetime influence maps}
\label{app:influence}

To gain qualitative insight into what spacetime regions of the
measurement record contribute to the neural-network prediction,
we compute influence maps that quantify the sensitivity of the
predicted entropy to local measurement events.

For a given spacetime location $(i,t)$, the influence is estimated
by measuring how strongly perturbations of the corresponding input
feature modify the predicted half-chain entropy $S_{N/2}$.
This produces a spacetime map that highlights the regions of the
measurement record that most strongly affect the model output.

Figure~\ref{fig:influence_maps} compares the resulting influence
patterns for representative architectures. The upper row shows the
influence maps for three different models at final time: a shallow single-scale
network with depth $K=2$, a deeper single-scale network with
$K=6$, and a hierarchical architecture RG$(2,2,2)$. In each panel,
the red cross marks the spacetime location associated with the
final-time readout near the entanglement cut.

Several qualitative features are apparent. The shallow network
exhibits a highly localized influence region, indicating that the
prediction depends primarily on measurements in a small neighborhood
of the readout location. Increasing the depth of the single-scale
network significantly enlarges the region of influence, reflecting
the increased spacetime scale accessible through message passing.
The hierarchical architecture produces an extended but coarser
influence region, consistent with its multiscale coarse-graining
structure.

The lower row illustrates the dependence of the influence region on
the readout time $t_r$. As the readout time increases, the dominant
influence shifts upward in time and broadens in spacetime, indicating
that later predictions depend on a larger portion of the preceding
measurement history. The resulting patterns resemble a causal
spacetime region surrounding the readout slice, qualitatively
consistent with the propagation of information in monitored
quantum circuits.

These influence maps provide an intuitive visualization of how the
graph neural network integrates information from the measurement
record and support the interpretation developed in the main text:
the architectural depth and hierarchy control the effective
spacetime scale over which the model aggregates information.

\bibliography{references/ref}
\end{document}